\documentstyle[12pt]{article}
\begin{document}
\def\rref#1{(\ref{#1})}
\title{ Hidden Higgs Particle}

\author{ S.G. Kovalenko\\
         Laboratory of Nuclear Problems,\\
	 Joint Institute for Nuclear Research,\\
         Head Post Office, P. O. Box 79, Moscow, RUSSIA}
\maketitle

\begin{abstract}
    A  modification  of  the  standard  model of electroweak
interactions with the  nonlocal Higgs  sector
is   proposed.   Proper    form   of   nonlocality    makes
Higgs particles unobservable after  the electroweak symmetry
breaking. They appear  only  as a virtual state   because their
propagator  is   an  entire   function.  We   discuss   some
specific consequences  of  this  approach  comparing it
with   the conventional standard model.
\end{abstract}

\section{Introduction}

     Many problems of the conventional standard model (SM) arise from
the presence of Higgs sector of scalar fields. Perhaps one
of the most evident problems is the absence of any
experimental manifestations of Higgs particles.  The ordinary explanation of
this fact by a large Higgs boson mass $M_H$ may be
unsatisfactory. At fairly large $M_H$, approximately at
$M_H\;\geq \; 1 \mbox{ TeV}$ ~\cite{Lee}, ~\cite{Chan}, the SM becomes
a strongly
interacting theory. In this case the usual perturbation theory (PT),
which is the only reliable method of calculation in quantum field
theory (QFT), cannot be applied to derivation of the SM predictions.
In the near future at the SSC and LHC the Higgs mass range
$M_H\;\leq\;800 \mbox{ GeV}$ will be exhaustively explored
 ~\cite{DiLel}, ~\cite{Altar}  and the above-mentioned
upper bound of the perturbative regime of the SM may be  exceeded.

    A situation when Higgses are not discovered is now under
serious theoretical consideration.
 The  composite models and schemes with dynamical mechanisms of gauge symmetry
breaking are investigated  extensively as an alternative of the models
with the fundamental scalars.  General properties of spontaneously
broken gauge symmetry
(SBGS) are also explored independently of concrete symmetry breaking
mechanisms ~\cite{Chan}, ~\cite{Dobad}. To our mind, it  is interesting to
consider  other approaches to SBGS without observable Higgs particles in
the ~framework of QFT.

   We propose an approach based on the idea of the so-called
"virton" field ~\cite{Efim} which can be constructed within the nonlocal
QFT. In the local one it doesn't exist. The main peculiarity  of the
 virton field is the following. After  quantization it describes not
ordinary particles, but unobservable quasi-particles, which appear  only
as virtual states. If we regard the virton fields as the Higgs
fields, we obtain an appropriate  model of the Higgs sector of SM which
generates the vector boson and fermion masses without  producing
observable Higgs particles.

   The nonlocal quantum field theory (NLQFT) is a self-consistent
scheme satisfying all principles of the conventional QFT (unitarity,
causality, relativistic invariance, etc.) and providing the basis for
correct description of the nonlocality effects. At the same time
the nature of nonlocality itself may be unknown. This point of view is
accepted in the series of work ~\cite{Moff} where a version of SM is
developed in which all interactions are nonlocal. There are some
problems with the gauge invariance because of nonlocality in gauge
field interactions. Nevertheless, this approach gives a good
possibility of building  completely ultraviolet finite theory of the
fundamental interactions. Another way to the finite QFT was
proposed in the work ~\cite{Krasnik} on the basis of infinite component
fields which also results in the introduction of the special form of
nonlocality.

    In our approach the nonlocality is introduced only in the Higgs
interactions. The main goal of this modification of the SM is to
exclude the scalar particles from the observable spectrum.
The theory in this case is certainly not finite, though its ultraviolet
property is improved and divergences of many diagrams are reduced.
Our method of quantization of the nonlocal fields also differs from the
one applied to this problem in the above mentioned approach
{}~\cite{Moff}. We use  intermediate regularization of the nonlocal
field theory by  introduction the ~infinite set of quantized local
auxiliary fields defined on the Hilbert space with the negative-norm
states.
\nopagebreak[4]
\section{Electroweak Symmetry Breaking and \- Nonlocal Self Interaction
          of Higgs Fields}

    We introduce  nonlocality into the Higgs self-interaction
writing down the Lagrangian of the scalar electroweak doublet
fields in the form
\begin{equation}
 -{\cal L} = {\phi}^{\dagger}(x)({\partial}^2 + m^2){\phi}(x) +
             \lambda  \left( {{\Phi (x)}^\dagger}*{\Phi (x)}
             \right)^2                                           \label{(1)}
\end{equation}
where $m^2 < 0 $ , and the nonlocal field $\Phi (x)$ is obtained
from the local one $\phi (x)$ by  "smearing" over the nonlocality
domain with the characteristic scale ${\ell}_0$. We don't
specify the nature of this nonlocality and, introducing
the phenomenological formfactor ${\cal K}$, define the nonlocal field
\begin{eqnarray}
 {\Phi (x)}  =  \int dy \ {\cal K}(x-y)\ \phi (y)
             =  {\cal K}\left({\ell}_{0}^{2} \ {\partial}^2\right)
             \ \phi (x)                                          \label{(2)}
\end{eqnarray}
   The nonlocal operator  ${\cal K}\left({\ell}_{0}^{2}\  {\partial}^2\right)$
can be presented in the form
\begin{eqnarray}
   {\cal K}\left({\ell}_{0}^{2}\ {\partial}^2\right)=  \sum_{n=0}^{\infty} \
   \frac{c_n}{(2n)!}\left({\ell_{0}^{2}}\ \partial^2\right)^2     \label{(3)}
\end{eqnarray}
  Then the generalized function  $  {\cal K}(x-y)=
{\cal K}\left({\ell}_{0}^{2}\ {\partial}^2\right)\delta (x-y)$
belongs to one of the spaces of nonlocal generalized functions which was
introduced and explored in the works  of Efimov ~\cite{Efim1}.

   Considering the theory based on the Lagrangian ~\rref{(1)} we follow
the method of quantization of the nonlocal fields developed in
these works.

    Let us rewrite the Lagrangian  ~\rref{(1)} in terms of the
nonlocal fields $\Phi(x)$
\begin{equation}
 -{\cal L}\  = \ {\Phi}^{\dagger}(x){{\cal K}^{-2}
                 \left({\ell}_{0}^{2}\ {\partial}^2\right)}
                 \left({\partial}^2 + m^2 \right){\Phi}(x)\  + \
                 \lambda  \left( {{\Phi (x)}^\dagger}*{\Phi (x)}
      	          \right)^2                                   \label{(4)}
\end{equation}
    We are looking for such conditions which, being applied to this
Lagrangian, guarantee the virton realization for the  scalar field
remaining after SBGS. In this case observable scalar particles will not
appear because their propagator  $G(p^{2})$ is an entire function.
Introduce a new operator
 \begin{equation}
    {\cal E}\left(\partial ^2 \right) =
    {\cal K}^{-2}\left({\ell}_{0}^{2}\ {\partial}^2\right)
    \left({\partial}^2 + m^2 \right) + \omega          \label{(5)}
\end{equation}
   The constant $\omega$ will be fixed from the condition
 $G(p^{2})={\cal E}^{-1}(-p^{2})$. Then we also require that the
function $\cal K$ is an entire analytic function without any zeros.
This means that the Higgs propagator $G(p^{2})$ has no poles after SBGS.
 The  ${\cal E}$-function must ~satisfy some conditions
following from the general principles of QFT. These are the Efimov
conditions ~\cite{Efim1}

\begin{itemize}
\item ${\cal E}(z)$ -is an entire analytic function of the order\-
      $ \frac{1}{2} \leq \rho \leq 1 $,                    \label{(aa)}
\item
${\left[{\cal E}(z)\right]}^{*} = {\cal E}(z^*) $          \label{bbb}
\item    ${\cal E}(z) > 0 $  for real  $ z $               \label{ccc}
\item In Euclidean momentum space ${\cal E}^{-1}({-p^2})$ has to
      decrease steeply enough for
 \begin{equation} \int\limits_{0}^{\infty}dp_{E}^{2}\
       {\cal E}^{-1}(p_{E}^{2}) <\infty  	\label{(6)}
\end{equation}
\end{itemize}
    The last condition results in decreasing Euclidean Green
functions of NLQFT. The most general form of the  ${\cal E}$-function
satisfying all these conditions is
\begin{equation}
    {\cal E}(z)\ = \ {\mu}^2 \exp \left({\cal W}(z)\right) \label{(EWz)}
\end{equation}
where $\mu$  is a parameter providing  ${\cal E}(z)$ with the correct
dimension ; $\cal W$ is a real entire  function increasing
with  $z^2\rightarrow \infty $.

 The interaction with the gauge fields is introduced by usual minimal
substitution
\begin{eqnarray}
	{\partial}_{\mu} \longrightarrow {\sl D}_{\mu}={\partial}_{\mu}-
        i g \frac{{\bf \tau}^a}{2}{\bf A}_{\mu}^a - i g'
	   \frac{Y}{2}\sf B_{\mu}                            \label{(7)}
     \end{eqnarray}
where  ${{\bf A}}_{\mu}$ and ${\sf B_{\mu}}$  are  the  $SU_{2L}$ and
$U_{1Y}$ gauge fields, respectively; Y is a weak hypercharge operator.

   Taking these into account, we rewrite the Lagrangian ~\rref{(4)}
in the form
\begin{equation}
 -{\cal L}\  = \ {\Phi}^{\dagger}(x)\left({\cal E}\left({\sl D} ^{2} \right)
                 \ - \ \omega \right) {\Phi}(x)\
                 + \ \lambda \left( {{\Phi (x)}^\dagger}*
                 {\Phi (x)}\right)^2                         \label{(8)}
\end{equation}

Gauge invariance of this Lagrangian is the direct consequence of the
fact that the  ${\cal E}(z)$-function is an entire one. Under
gauge transformations
\begin{eqnarray}
\Phi \  &\stackrel{g}{\longrightarrow}& \ {\Phi}^g=g \Phi\\    \label{(9)}
{\sl D}_{\mu}  &\stackrel{g}{\longrightarrow}& \
{\sl D}_{\mu}^g=g{\sl D}_{\mu}g{-1}               \label{(10)}
 \end{eqnarray}
the operator ${\cal E}({\sl D} ^{2}) $, being the sum of positive
 degrees of the covariant derivative $\sl D$, transforms as
\begin{equation}
      {\cal E} \stackrel{g}{\longrightarrow}
      {\cal E}^{g} = {\cal E}(g{\sl D}^{2}g^{-1} )=
       g{\cal E}({\sl D}^{2}) g^{-1}                   \label{(11)}
\end{equation}

   Noteworthy is that  according to the Picard theorem on $\bf a$ - points
of an entire function ~\cite{Titch},  the ${\cal E}(z)$ function takes the
$\omega$-value an infinite number of times. This statement can be easily
understood
with the simplest example of function $\exp{z}$.
 Consequently, the propagator  $\left({\cal
E}(-{p}^{2})-{\omega}\right)^{-1}$ of the $\Phi$ -- field has an infinite
number of poles, some of them at negative or imaginary values of  $p^2$.
In quantum theory they correspond to particles with unphysical complex
masses. As is well known, the presence of these states,  analogous to
"tachyon" states in the conventional SM, are the signal of SBGS.

   Let us consider the quantization problem for the theory based on the
Lagrangian ~\rref{(8)}. The standard canonical procedure cannot be
directly  applied in this case due to the presence of higher time
derivatives in the kinetic term. Therefore, following ~\cite{Efim1}, we
consider the properly regularized theory
 \begin{eqnarray}
	       -{\cal L}^{\delta}\                       \label{(12)}
       = \ {\Phi}^{\delta \dagger}(x)
         {\cal  E}^{\delta} \left({\sl D} ^{2} \right)
         {\Phi}^{\delta} (x)\ - \ \omega^{\delta}{\Phi}^{\delta \dagger}(x)
         *{\Phi}^{\delta} (x)\ + \\
         +  \ \lambda \left(
         {\Phi}^{\delta \dagger}(x) *{\Phi}^{\delta} (x)\right)^2
\end{eqnarray}
where $\delta$ is a regularization parameter. The regularization is
chosen in  such a way that
\begin{equation}
\lim_{\delta \rightarrow 0}                              \label{(13)}
{\cal  E}^{\delta} \left({\sl D} ^{2} \right)
 = {\cal  E}\left({\sl D} ^{2} \right)
\end{equation}
and the regularized function

\begin{equation}
{\cal  E}^{\delta} \left(p^{2} \right) \sim \prod_{j=1}^{\infty}
\left(p^2 - m_{j}^{2}(\delta)\right)                      \label{(14)}
\end{equation}
has an infinite number of zeros in a sequence of points such that
 $ m_{j}^{2} (\delta)>0 $ and $ m_{j}^{2} (\delta) \rightarrow \infty $
when $ \delta \rightarrow 0 $.

   Then the theory can be quantized on the Hilbert space with the
negative-norm states \cite{Pais}.

   For  the   $\delta  >  0$ the  free  Hamiltonian  $H_{0}^{\delta}$,
 the $  S^{\delta} $--matrix, the Green functions   $ G^{\delta}  $
and other objects of QFT can be constructed. It is accepted
{}~\cite{Efim1} that in the limit  $ \delta \rightarrow 0 $ this
construction gives the solution of a quantization problem for the
initial nonlocal system with the Lagrangian ~\rref{(12)}.

  The first step is to represent the meromorphic function
$ \left[{\cal  E}^{\delta}\right]^{-1} $  in the form

\begin{equation}
     \left[{\cal  E}^{\delta}(p^2)\right]^{-1}=            \label{(15)}
     \sum_{j=0}^{\infty}(-1)^{j} \frac{{\cal A}_{j}(\delta)}
     {m_{j}^{2}(\delta) - p^2}
\end{equation}
 where
\begin{equation}
     m_{j}^{2}(\delta) =  \frac{(j+1)^{\sigma}}{\delta}M_{0}^{2}  \label{(16)}
\end{equation}
  and $ \sigma< 1/{\rho} \leq2 $ , $ M_0 = l_{0}^{-1}$.
  The coefficients  ${\cal A}_{j}(\delta)$ can be easily calculated for
  any concrete function.
Define the infinite set of Pais -- Uhlenbeck auxiliary fields

\begin{eqnarray}
       {\Phi}_{j}^{\delta}(x)\ &=&\ \sqrt{{\cal A}_{j}(\delta)}
    \    \frac{{\cal  E}^{\delta} \left({\sl D} ^{2} \right)}
       {{\sl D} ^{2} + m_{j}^{2}(\delta)}                 \label{(17)}
      \  {\Phi}^{\delta}(x) \\
	  {\Phi}^{\delta} (x)\ &=&\ \sum_{j=0}^{\infty}(-1)^{j}
       \ \sqrt{{\cal A}_{j}(\delta)}
         \ {\Phi}_{j}^{\delta}(x)                          \label{(18)}
\end{eqnarray}
	In terms of the fields  $ {\Phi}_{i}^{\delta} $
the regularized Lagrangian  ~\rref{(12)} reads
\begin{eqnarray}
       -{\cal L}^{\delta}\                       \label{(19)}
       = \ \sum_{j=0}^{\infty}{(-1)^j}\ {\Phi}_{j}^{\delta \dagger}(x)
         \left({\sl D} ^{2}\ + \  m_{j}^{2}(\delta)\right)
         {\Phi}_{j}^{\delta} (x)\ -  \nonumber\\
         - \ \omega^{\delta} {\Phi}^{\delta \dagger}(x)
         *{\Phi}^{\delta} (x)\ + \ \lambda \left(
         {\Phi}^{\delta \dagger}(x) *{\Phi}^{\delta} (x)\right)^2
\end{eqnarray}
     Considering  $ {\Phi}_{i}^{\delta} $ to be independent
fields we can quantize the theory with this Lagrangian  in the framework
of canonical formalism. Then the equal-time canonical commutation
  relations  take the form

\begin{equation}
	 \left[{\Phi}_{i}^{\delta \dagger}(0,\vec{x}),\
	 {\Phi}_{j}^{\delta} (0,\vec{y})\right]\ =\ i(-1)^{j}
         {\delta}_{ij} {\delta}                  \label{(20)}
\left(\vec{x}-\vec{y}\right) \end{equation}
 Due to the presence of the factor $(-1)^j $ these relations can be realized
only on the Hilbert space with  the negative-norm quantum states.
 Moreover, it can be shown ~\cite{Efim1}
that on this space the quantum field ${\Phi}^{\delta}$ is a local
operator.

  Considering the SBGS    we make standard shift of all fields
${\Phi}_{j}^{\delta} $ independently

\begin{eqnarray}
	  {\Phi}_{j}^{\delta} (x)\ &=& \         \label{(21)}
\left(\eta_{j}^{\delta}(x)\ + \ v_{j}^{\delta}
\right)\frac{\chi}{\sqrt{2}}\\
	  {\Phi}^{\delta} (x)\ &=& \               \label{(22)}
\left(\eta^{\delta}(x)\ + \ v^{\delta}
\right)\frac{\chi}{\sqrt{2}}\\
\chi\ &=& \ \left(0 \atop 1 \right)   \nonumber
\end{eqnarray}
 where
\begin{eqnarray}
    {\eta}^{\delta} (x)\ &=&\ \sum_{j=0}^
       {\infty}(-1)^{j}                         \label{(23)}
       \ \sqrt{{\cal A}_{j}(\delta)}
        \ {\eta}_{j}^{\delta}(x)\\
    {v}^{\delta} (x)\ &=&\ \sum_{j=0}^{\infty}(-1)^{j}
       \ \sqrt{{\cal A}_{j}(\delta)}
         \ {v}_{j}^{\delta}(x)                   \label{(24)}
\end{eqnarray}
     Redefining the theory in terms of the fields
$\eta_{j}^{\delta}$ we fix the values of $v_{i}^{\delta}$ and
$\omega^{\delta}$ parameters from the   requirement of  the absence in
the resulting
Lagrangian of any terms which are  linear or quadratic  in fields
$\eta_{j}^{\delta}$ ~except $m_{i}^{2}(\delta){\eta_{i}^{\delta}}^{2}$.
Then we obtain the equations

\begin{eqnarray}
    & &m_{i}^{2}(\delta)v_{i}^{\delta}\ - \ v^{\delta}
\sqrt{{\cal A}_{j}(\delta)} \left(\omega^{\delta} - {\lambda}
\left(v^{\delta} \right)^2 \right) = 0\\         \label{(25)}
& &\omega^{\delta}\  = \ 3 \lambda \left(v^\delta \right)^2   \label{(26)}
\end{eqnarray}
   The solution of these equations reads
\begin{eqnarray}
\omega^{\delta} \ &=& \ \frac{3}{2} \  {\cal  E}^{\delta}(0) \\ \label{(27)}
\left(v^{\delta}\right)^2\ &=& \ \frac{{\cal  E}^
{\delta}(0)}{2\ \lambda}\\                           \label{(28)}
\left(v_{i}^{\delta}\right)^2 \ &=& \
\frac{{\cal A}_{j}(\delta)\left({\cal  E}^
{\delta}(0)\right)^3}{2\ \lambda \ m_{i}^{4}(\delta)}           \label{(29)}
\end{eqnarray}

   Taking this into account and substituting ~\rref{(23)} - ~\rref{(24)}
to ~\rref{(21)}  one finds after some transformations the final
Lagrangian  in the unitary physical gauge

\begin{eqnarray}
      &-&{\cal L}^{\delta}\                               \label{(30)}
      = \frac{1}{2}\ \sum_{j=0}^{\infty}{(-1)^j}\
         {\eta}_{j}^{\delta}\left({\partial}^{2}\ + \
         m_{j}^{2}(\delta)\right) {\eta}_{j}^{\delta} \ + \
          \lambda  v^{\delta}
          \left({\eta}^{\delta}\right)^3 \ + \
           \frac{\lambda}{4}\left({\eta}^{\delta}
           \right)^4 \ -   \nonumber      \\
	  &-& \frac{g^2}{4}
	   \sum_{j=0}^{\infty}{(-1)^j}\ \left[ \left({\eta}_{j}^
           {\delta} \right)^2 \ + \ 2 \eta_{j}^{\delta}
	     v_{j}^{\delta} \right]\left( W_{\mu}^{+}
             W^{- \mu} \ + \ \frac{1}{2\cos^2 {\theta_{W}}}
             Z_{\mu} Z^{\mu} \right)\ - \nonumber \\
	 &-&\ \ \ \ M_{W}^{2}\ W_{\mu}^{+} W^{- \mu} \ - \
	     \frac{1}{2}M_{Z}^{2}\  Z_{\mu}Z^{\mu} \ - \nonumber \\
          &-&\ \ \ \ \sum_{f}^{} {\bar{\psi}}_{f}\left(i{\gamma}_{\mu}
	     {\sl D}^{\mu} \ - \ m_{f} \left(1+\frac{\eta^
             {\delta}}{v^{\delta}}\right)\right)\psi_f
\end{eqnarray}
    For the completeness the fermion part of the model is also
included in this expression. The masses of the W and Z -- bosons are
given   by

\begin{eqnarray}
 M_{W}^{2} \ &=& \ \frac{g^2 v^2\ {\cal E'}(0)}{4}\ =\    \label{(31)}
  \frac{g^2   \ {\cal E}(0){\cal E'}(0)}{8\ \lambda}\\
	  M_{Z}^{2}\ &=& \  \frac{M_{W}^{2}}{\cos^2
{\theta_{W}}}                             \label{(32)}
\end{eqnarray}
     The Lagrangian ~\rref{(30)} gives the solution of the problem
we formulated at the beginning.
In the limit  $ \delta \rightarrow 0 $ it describes the SM with
a nonlocal Higgs sector.š The scalar particles corresponding to the quantized
fields ${\eta}_{j}^{\delta}(x)$  escape from the observable spectrum
for their masses $m_{j}^{2}(\delta)\rightarrow \infty$
when  $ \delta \rightarrow 0 $. Moreover, the kinetic terms of these
fields in  the Lagrangian ~\rref{(30)}  can be written in the form

\begin{equation}
	 \sum_{j=0}^{\infty}{(-1)^j}\ {\eta}_{j}^{\delta}(x)
	 \left({\partial}^{2}\ + \  m_{j}^{2}(\delta)\right)
	 {\eta}_{j}^{\delta}(x)\ = \ \eta^{\delta}(x)     \label{(33)}
	 {\cal  E}^{\delta} \left({\partial}^{2} \right)
         \eta^{\delta}(x)
\end{equation}
Thus in the limit $ \lim_{\delta \rightarrow 0} \eta^{\delta}\ =\ \eta $
one obtains the field  $\eta$  with the corresponding propagator
$ {\cal  E}^{-1}(p^2)$ is being an entire function.  This is another
condition of absence of the observable scalar particles.
    From formulae ~\rref{(30)}, ~\rref{(32)} it can be deduced
that  our model differs from  the conventional SM only within the  Higgs
sector and its interactions with the gauge fields. The mass formulae for the
$W$ and $Z$-- bosons slightly differ but this difference is not
essential as it will be shown further.

   The calculations of the S-matrix elements in our model  mainly
are based on the standard technique of QFT perturbation theory.

   The Feynman rules can be extracted directly from the
Lagrangian ~\rref{(30)}. Here we write down only the
${\eta}_{j}^{\delta}(x)$ -- field propagators:

\begin{eqnarray}
  G_{ij(\bf c)}^{\delta}(x-y)\ &=& \ i<0|\ T\left(\eta_{i}^
   {\delta}(x)\ \eta_{j}^
   {\delta}(y)\right)|0>\ = \nonumber \\
   &=&\ (-1)^j\  \delta_{ij}\
   {\cal D}_{j(\bf c)}^{\delta}(x-y)  \\   \label{(34)}
\nonumber \\
\mbox{where \ \ \ \ \ \ \ \ \ \ \ \ \ \ \ \ \ \ \ }&& \nonumber \\
\nonumber \\
{\cal D}_{j{\bf c}}^{\delta}(x)\ &=& \           \label{(35)}
     \int \frac{d^4k}{(2\pi)^4}\
    \frac{e^{-ikx}}{m_{j}^{2}(\delta) - k^2 - i\epsilon}
\end{eqnarray}
   is the propagator of the local scalar field of the mass
$m_{j}^{2}(\delta)$.

The essential peculiarities of the calculations in the framework of
our nonlocal modification of the minimal SM are as follows. The diagrams
with the external  ${\eta}_{j}^{\delta}$-lines must be excluded,  but
all possible internal ${\eta}_{j}^{\delta}$-states must be summed up.
Final formulae for the physical matrix elements  are obtained in the limit
$ \delta \rightarrow 0 $.

   It is relevant to note that the diagrams with the internal
${\eta}_{j}^{\delta}$ -- lines are less divergent in comparison with the
conventional SM. In some cases these diagrams are found to be
 ultraviolet finite. This is because  of the fast decrease of the
Euclidean Green functions in the NLQFT. In the present paper this
property is expressed by the condition \rref{(6)}.   If all
the interactions are  supposed to be nonlocal as accepted in the
above-mentioned  approach of Moffat  ~\cite{Moff} , then the theory becomes
finite.

\section {Model Parameters and Domain of Perturbative Regime}

   To investigate the main features of the proposed nonlocal modification of
the SM let us consider the simplest "minimal" variant of nonlocality.
It corresponds to the following choice ${\cal W}(z)=z/M_{0}^{2}$
in formula ~\rref{(EWz)}. This gives

\begin{equation}
  {\cal E}(z)\ = \ \mu^2 \exp( z/M_{0}^{2})       \label{(Min)}
\end{equation}
where  $\mu$ and $M_0$ are free parameters, extracted
from the experiment. Regarding their possible values one notes that
the  $\mu$-parameter has no essential physical meaning. The proper
$\eta$-fields and the $\lambda$-coupling constant redefinition
change it to any arbitrary value.  In fact, the Lagrangian  ~\rref{(30)}
is invariant under the following transformation

\begin{eqnarray}
\eta^{\delta}\ &\longrightarrow& \ {\eta'}^{\delta}\ =\
{\kappa}^{-\frac{1}{2}} \
\eta^\delta\\                                     \label{(Scal1)}
\eta_{i}^{\delta}\ &\longrightarrow& \ {\eta'}_{i}^{\delta}\ =\
\eta_{i}^{\delta}\\                       \label{(Scal2)}
\lambda\ &\longrightarrow& \ \lambda'\ =
\ {\kappa}^2 \ \lambda\\                     \label{(Scal3)}
{\cal E}^{\delta}(\partial^2)\ &\longrightarrow& \  {\cal E'}^{\delta}
(\partial^2) \ =\ {\kappa}\ {\cal E}^{\delta}(\partial^2)      \label{(Scal4)}
\end{eqnarray}
The latter transformation results in
$\mu^2 \longrightarrow {\mu'}^2=\kappa \mu^2$.
In the considered modification of the SM there are no
asymptotic conditions on the normalization of free fields  for the
absence of observable scalar particles. Therefore the physics is
independent of the scale factor  $ \kappa $ and value of  $ \mu^2 $.
The only role of the $\mu$ -- parameter is to provide a correct
dimension of  the  propagator  of the scalar  field  $\eta$.
For the simplicity we take  $\mu=M_0$. Then the mass formulae
{}~\rref{(31)}--~\rref{(32)} acquire the same form as in the conventional
SM.

  For the practical calculations in the framework of  our
approach it is important to know the limitations  on
applicability of perturbation theory. These limitations come
from the partial wave unitarity ~\cite{Dicus}, ~\cite{Lee} which
require

\begin{equation}
\left|{a_J}(s)\right|\ \leq \ 1       \label{(AJ)}
\end{equation}
To find these limitations it is enough in our case to calculate the
zeroth partial wave amplitude $a_0$   for longitudinal $W$ and $Z$ -- boson
scattering $W_{L}W_{L}\;\longrightarrow\;Z_{L} Z_{L} $. For the
Lagrangian ~\rref{(30)} with the  ${\cal E}$-function in the
form ~\rref{(Min)} we have in tree approximation

\begin{equation}
a_0\ = \ \frac{g^2}{64\pi}\
\frac{M_{0}^{2}}{M_{W}^{2}}\ \left(
 e^{S/M_{0}^{2}}  -\ 1 \right)     \label{(A0)}
\end{equation}
 this gives the limitation
\begin{equation}
S\ \leq \ S_{max} (M_0)\ = \ M_{0}^{2} \                 \label{(Smax)}
ln \left(\frac{8\pi\sqrt{2}}{G_F M_{0}^{2}}\ +\
1\right)
\end{equation}
 In this kinematic domain the condition
$\left|{a_0}(s)\right|\;\leq\;1$ is satisfied. The function
$ S_{max} (M_0)$  monotonically increases with  $M_0$ and asymptotically
tends to a constant

\begin{equation}
  \lim_{M_0 \rightarrow\infty} S_{max} ({M_0}) \ = \
  \frac{8 \sqrt{2} \pi}{G_F}\ \simeq \
 {\left(1.8\ TeV\right)}^2                          \label{(Smax1)}
\end{equation}
The saturation occurs rather quickly, and  for  $M_0\;=\;1.5 \mbox{ TeV}$ we
get the upper bound $\sqrt{S}\leq\sqrt{S_{max}}\simeq1.5 \mbox{ TeV}$.
To achieve higher energies it is necessary to perform  the calculations
in the next to leading order of perturbation theory and maybe to sum up some
classes of diagrams.

   Note that the obtained limitation ~\rref{(Smax)} strongly
depends on the form of the ${\cal E}$-function. Starting from the
function differing from  ~\rref{(Min)} we arrive at the
limitation other than ~\rref{(Smax)}.

     In the conventional local SM the unitarity gives an upper bound on the
Higgs boson mass. At the tree level this is $M_H\leq 2 \mbox{ TeV}$.  If the
Higgs boson is not discovered in this mass region, then
perturbation theory is not applicable for the derivation of the SM
predictions at high energy. Moreover, some of the low energy
predictions obtained within this framework are subject to doubt.
The reason is as follows. Because in the SM there is a well known relation

\begin{equation}
M_{H}^{2}\ =\ 2\lambda v_{SM}^{2}         \label{(Mlamb)}
\end{equation}
where $v_{SM}\ \simeq \ 250 \mbox{ GeV}$, then for $M_H\;>\;1 \mbox{ TeV}$ we
have $\frac{\lambda}{4\pi^2}\;>\;1$. The last quantity
characterizes perturbative corrections due to the Higgs
self-interaction. Therefore, if it is not small enough, the
perturbation theory fails. This conclusion does not depend on
the kinematic domain considered but only on the $M_H$ -- value.
As we have seen, in the proposed nonlocal modification of SM  the
situation is different. There is no connection between
Higgs invisibility and  applicability of the perturbation
theory. It has been shown that the lowest order calculations are correct
in the definite kinematic domain  which
depends on the form of nonlocality in the Higgs
self-interaction.  For the minimal variant  ~\rref{(Min)} the domain is
defined by   ~\rref{(Smax)}  .

\section{On Physical Meaning of the $\eta$-Field}

    We can futher clarify the physical meaning of the nonlocal
$\eta$-field and the parameter $M_{0}$ of the minimal
ansatz \rref{(Min)}, considering the static limit for two heavy sources
$J_{i}$ interacting via  $\eta$-quantum exchange.

The interaction can be described by the Lagrangian

\begin{equation}
{\cal L} = \kappa \left[ J_{1} \eta +
          J_{ 2} \eta + h.c. \right]         \label{(Lint)}
\end{equation}

In the lowest order of the perturbation theory
the interaction energy (potential) of two sources
takes the form:
\begin{equation}
W = \frac{\kappa^{ 2}}{2} \int
     dx_{1}dx_{2} \left[
     J_{1}^{\dag}(x_{1}) G(x_{1} - x_{2})J_{2}(x_{2})
     + h.c.\right]                            \label{(Energ)}
\end{equation}

where $G(x)$ is a propagator of the nonlocal $\eta$-field.

Considering in the static limit the sources $J_{1,2} $
motionless and located at points $\vec r_{1,2}$ respectively
we can write down
\begin{equation}
J_{ i}(x) = J_{ i}({\vec x}, t) =
               \rho \ \delta (\vec x - \vec r_{ i})    \label{(Rho)}
\end{equation}

with the normalization condition
\begin{equation}
{\rho}^{\ast} \rho = 1          \label{(Norm)}
\end{equation}

For the minimal ansatz in this limit we can easily obtain
\begin{equation}
W(r) = const \  e^{-r^{ 2}M_{ 0}^{ 2}/4}   \label{(Pot)}
\end{equation}

Here $r = \mid \vec r_{ 1} -  \vec r_{ 2} \mid$.

The potential ~\rref{(Pot)} falls down faster than the Yukawa potential
\begin{equation}
W_{ Yuk}(r) = const \  \frac{e^{-rm}}{r}   \label{(Yuk)}
\end{equation}

induced by the exchange with a particle of mass $m$.

The nonlocal character of the $\eta$-induced interaction in this case
manifests itself by the absence of the short-distance
$r\rightarrow 0$ singularity. The latter is a genuine
property of the local interactions to which the Yukawa
interaction belongs.

To make clearer the physical role of the $M_{0}$-parameter
let us calculate the mean square radius of the "$\eta$-cloud"
around the point-like source.  It is defined by the potential $W(r)$
as follows
\begin{equation}
<r^{ 2}>_{ \eta} = \int dr r^{ 2} W(r)/\int dr W(r) \ = \
\frac{6}{M_{ 0}^{ 2}}                \label{(r2)}
\end{equation}

Similar quantity for the Yukawa potential is
\begin{equation}
<r^{2}>_{Yuk} \ = \  \frac{6}{m^{2}}                \label{(r22)}
\end{equation}

On comparing ~\rref{(r2)}  with ~\rref{(r22)} we can interpret
the $M_{ 0}$-parameter as an "effective"
mass of the hidden $\eta$-particles.
\section{Conclusion}

    Thus, the introduction of  nonlocality in the interactions
of Higgs fields
enables us to exclude the scalar Higgs particles from
the  observable spectrum of the SM. We don't specify the nature of this
nonlocality. In principle, it may be considered not  a fundamental
physical notion but  an effective phenomenological way of
taking into account some of interactions  beyond the SM . Proposed
nonlocal modification of the SM is described by the renormalizable
Lagrangian   ~\rref{(30)} for which we have formulated the rules of
perturbative calculations of physical matrix elements. Though the
applicability of the lowest order calculations is bounded within the
energy domain ~\rref{(Smax)}  in this framework one can, nevertheless,
calculate most of the electroweak effects and compare the predictions
with the experimental data. These will be considered elsewhere in  our
next paper. Here we confine ourselves to the following general
remarks. The main physical difference between the proposed nonlocal
modification of the SM and the conventional one comes from their Higgs
sectors. However, the Higgs field interacts with the leptons and
quarks, except the heavy t-quark, rather weakly. Thus we may expect the
observable difference in predictions of these two variants of the SM at
the level of refined effects of the radiative corrections or in the
processes difficult for the experimental investigation. Among them there
are widely discussed processes  of the $W$ and $Z$ -- boson scattering.
They  will be  accessible in the near future  at the SSC and LHC --
colliders where it is planned   to search for possible growth with the
energy of their cross sections. This behaviour is predicted in the
conventional SM for the longitudinal vector boson scattering, for
example, $W_{L} W_{L}\;\longrightarrow \; Z_{L} Z_{L}$. While the
nonlocal modification predicts the growth for the every polarizations of
$W$ and $Z$ -- bosons. Another discrepancy is the radiative
corrections. They also depend on the nature of the Higgs sector. One can
expect the most considerable  difference in the predictions of
the nonlocal modification and the conventional SM for those radiative
effects  which, being calculated in the conventional SM, have the
strongest dependence on the ~Higgs boson mass.

\end{document}